# Anecdotal Survey of Variations in Path Stroking among Real-world Implementations

Mark J. Kilgard
NVIDIA

July 23, 2020

**Abstract**

Stroking a path is one of the two basic rendering operations in vector graphics standards (e.g., PostScript, PDF, SVG). We survey path stroking rendering results from real-world software implementations of path stroking for anecdotal evidence that such implementations are prone to rendering variances. While our survey is limited and informal, the rendering results we gathered indicate widespread rendering variations for simple-but-problematic stroked paths first identified decades ago. We conclude that creators of vector graphics content would benefit from a mathematically grounded standardization for how a stroked path should be rasterized.

## 1 Introduction

Vector graphics standards such as PDF specified by [2], Scalable Vector Graphics (SVG) specified by the [8], PostScript specified by [1], HTML5 Canvas specified by [9], PCL specified by [5], and XPS specified by [4] all specify two basic rendering operations on paths: filling and stroking.

### 1.1 Lack of Mathematical Grounding for Path Stroking

While all these standards define filling a path in terms of a pixel's winding number with respect to the path as computed by a closed contour integral, none of these standards provides a rigorous, mathematically grounded description of precisely what it means to stroke a path.

For example, the PDF standard states its stroke operator "shall paint a line along the current path" and "shall follow each straight or curved segment in the path, centered on the segment with sides parallel to it." Other standards are no more rigorous. The PDF standard's description is intuitive in its appeal to a painting metaphor. However a metaphor is insufficient to reason about what pixels should and should not be covered by any particular stroked path segment.

For this reason, we survey real-world software implementing path stroking to study what variations in stroke rasterization exist. We anecdotally examine path stroking rendering results from 20 different software implementations and report our findings.

## 1.2  Caveats and Disclaimers to Our Survey

While we believe our survey is informative, we make *no* claims that our survey is comprehensive or scientific. Our software selection process was essentially the vector graphics software that we had readily available to us. That said, we also made no effort to exclude vector graphics software available to us; all the results we collected are included in our survey.

These results are not intended to "name and shame" any software implementation. Indeed, as the rasterization of stroked paths is not rigorously specified, there is no conformance standard to base such a judgment. We suspect some of the reported results are simply implementation bugs. We made no attempt to determine what is a bug, what is a justified performance-correctness trade-off, or what might be an alternative plausibly correct path stroking rasterization result.

We did not conduct a user study; the presented judgments are our own.

We performed no sensitivity analysis to determine when precisely rendering variances emerge.

We justify all these caveats and disclaimers because we are focused on making broad determinations about 1) whether or not rendering variances in path stroking exist in software used today, and 2) what, if any, evidence exists for a consensus about how stroked paths should be rasterized.

## 2  Test Cases

We evaluated just four test cases. Each test case is a simple path consisting of a single cubic Bézier curve that either exactly forms a cusp or nearly does. The stroke width is wide relative to the arc length of the curved segment. Each stroked path is drawn without end caps. All the control point coordinates are exact integer values.

An article by Corthout and Pol describing the design of a chip for rasterizing PostScript [3] is the source of our four test cases. This paper contains a figure consisting of a fragment of PostScript code (quoted verbatim below) and its Adobe PostScript (circa 1991) rendering results.

## 2.1 PostScript Version

Corthout and Pol provide this PostScript fragment implementing their test cases:

```
50 setlinewidth
0 0 moveto
110 100 -10 100 100 0 curveto stroke
150 0 translate
0 0 moveto
101 100 -1 100 100 0 curveto stroke
150 0 translate
0 0 moveto
100 100 0 100 100 0 curveto stroke
150 0 translate
0 50 moveto
10 60 0 60 10 50 curveto stroke
showpage
```

We graphed the generator curves for these test cases:

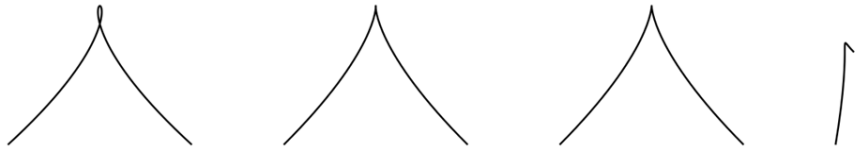

As you can observe, the first (from left-to-right) test case is nearly a cusp but forms a small loop; the second forms a minuscule loop; the third forms an exact cusp; and the fourth is a serpentine cubic Bézier segment close to being a cusp and with its right end point near the forming cusp.

The image below reproduces the figure from Corthout and Pol showing what PostScript (circa 1991) renders for their test cases:

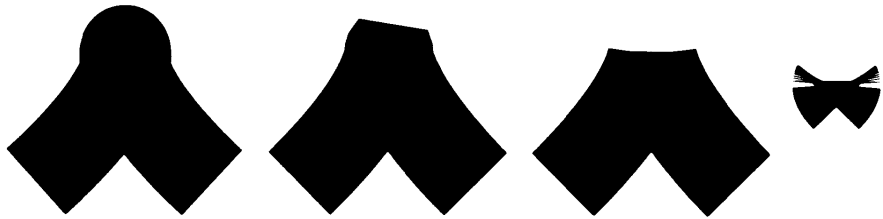

## 2.2 SVG Version

We converted their test cases into SVG for our testing:

```
<?xml version="1.0" standalone="no"?>
<!DOCTYPE svg PUBLIC "-//W3C//DTD SVG 1.1//EN"
  "http://www.w3.org/Graphics/SVG/1.1/DTD/svg11.dtd">
<svg width="800px" height="800px" viewBox="-100 0 1600 1600"
     xmlns="http://www.w3.org/2000/svg" version="1.1">
  <g transform="translate(0, 400) scale(2,-2)"
     stroke-width="50" stroke="black" fill="none" >
    <path
      d="M 0 0 C 110 100 -10 100 100 0 "/>
    <path transform="translate(150,0)"
      d="M 0 0 C 101 100 -1 100 100 0 "/>
    <path transform="translate(300,0)"
      d="M 0 0 C 100 100 0 100 100 0 "/>
    <path transform="translate(450,0)"
      d="M 0 0 C 10 60 0 60 10 50 "/>
  </g>
</svg>
```

Once in SVG, we were able to convert automatically the test cases into a number of standard vector graphics formats.

To encourage reproducing or expanding on our survey results, we make available six test files named: `PathStrokingSurvey.{svg,pdf,eps,emf,dxf,tex}`

## 3 Methodology

### 3.1 Collecting Rendering Results

We did not focus on testing the "latest and greatest" version of a particular software implementation; we tested what we had—though often that was a fairly current version of the software, particularly for auto-updating software such as web browsers. We provide version information we believe sufficiently detailed to replicate our rendering results. Our rendering results were collected on Windows-base PCs with Intel-based CPUs and NVIDIA-based GPUs. When we had a choice of rendering pixel resolution, we attempted to render the four test cases into an image roughly 1000 pixels wide.

Not every software implementation supported loading SVG content. When SVG import was not available we favored PDF, otherwise Encapsulated PostScript, otherwise Windows Enhanced Metafile (EMF). We used Inkscape to export our non-SVG alternatives. When SVG was not available, we did not observe any situations where different supported import formats affected the rendering results. We used a screen capture program to grab the actual pixel's displayed to the user.

## 3.2 Grading Rendering Results

To grade the results, we retrospectively decided on a qualitative grading system. We judged a test case "satisfactory" based on four criteria:

- if the loop or cusp (for the first 3 cases) is well-formed, rounded, and without obvious faceting or pixelation;

- if the stroked path is free from internal pixel-scale holes;

- if the fourth (rightmost) case has a notch without that notch being faceted or obviously circular; and

- if the stroke is fully contained within the dilation of the cubic Bézier segments control cage by the stroke radius.

These criteria are consistent with a model of stroking whereby an idealized "wide but thin" pen nib is centered on the path and orthogonal to it and then carefully traces the path. At an ideal cusp, the nib rotates 180 degrees at the cusp before continuing along the path past the cusp. The width of the nib is the width of the stroke. Those pixels swept out by the nib, plus any additional pixel's covered by the path's caps and joins, are part of the path's stroked region.

When all four criteria are present, we rate the grade an "**A**"; missing one criteria is graded "**B**"; missing two, a "**C**", and missing all four, a "**D**". While somewhat arbitrary, subjective, and developed retrospectively, we believe this grading scale is sufficiently unambiguous you will concur with our grading.

In borderline situations, we settle for a minus rating (e.g., "**B−**"). We reserve "**A+**" for implementations that both satisfy all four criteria and match a consensus of other implementations also meeting all four criteria.

By this grading system, Corthout and Pol's PostScript rendering result in Section 2.1 is graded **C** as only the leftmost test case satisfies all the criteria above.

## 4 Rendering Results

We collected 20 rendering results. We list our rendering results in alphabetical order by software implementation name. When we were aware a software implementation had distinct path stroking implementations (e.g., Direct2D and Skia), we attempted to collect the distinct results and report them as such.

### 4.1 Acrobat Reader DC 2019

Details: Adobe Acrobat Reader DC 2019.012.

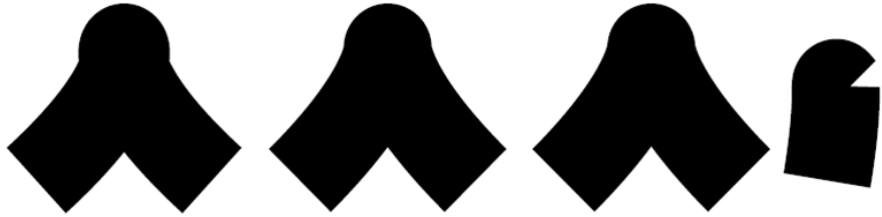

Grade: **A+**.

### 4.2 Cairo 1.12.2

Details: Cairo 1.12.2, circa 2012.

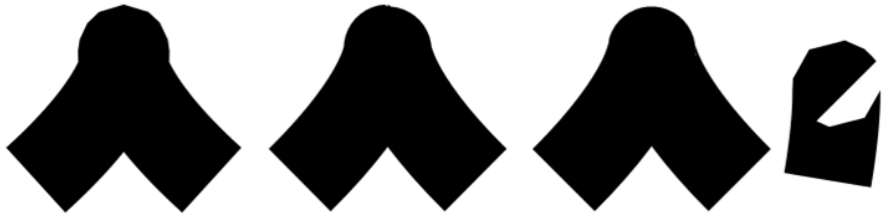

Grade: **B−**, faceted notch on rightmost path.

### 4.3 Chrome 74

Details: Chrome 74.0.3729, released 2019.

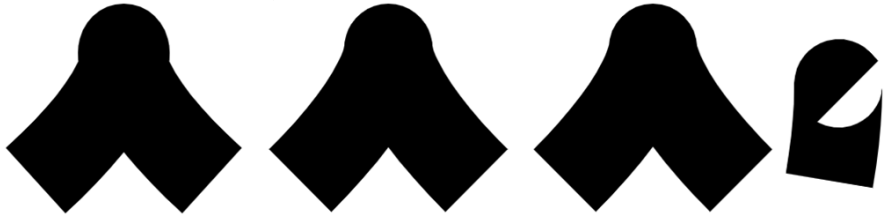

Grade: **B**, rounded notch on rightmost path.

### 4.4 Direct2D

#### 4.4.1 Direct2D GPU

Details: Direct2D GPU 1.1, NVIDIA GPU, 2019.

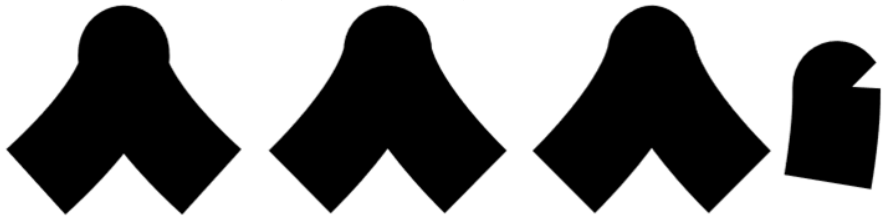

Grade: **A+**.

### 4.4.2 Direct2D CPU

Direct2D 1.1 also provides a WARP (Windows Advanced Rasterization Platform) renderer described as a "high speed, fully conformant software rasterizer."

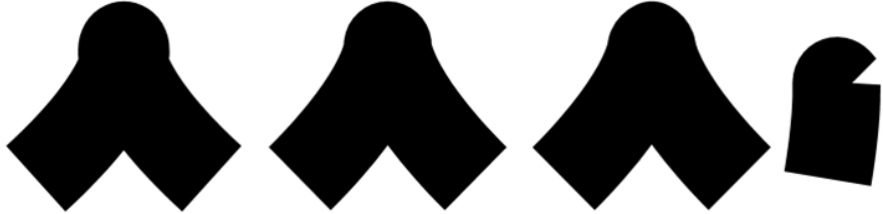

Grade: **A+**.

## 4.5 Firefox 66

Details: Firefox 66.05, released 2019. Likely using Direct2D.

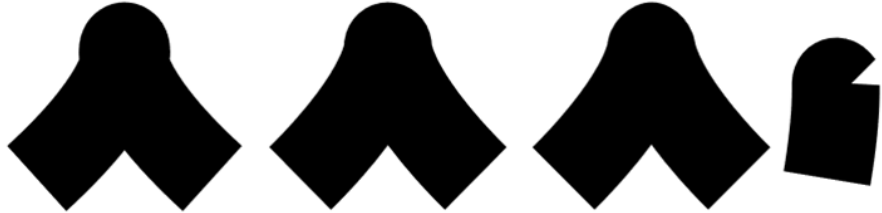

Grade: **A+**, likely due to Firefox rendering with Direct2D.

## 4.6 Foxit Reader 8

Details: Foxit Reader 8.0.2.805, released 2016.

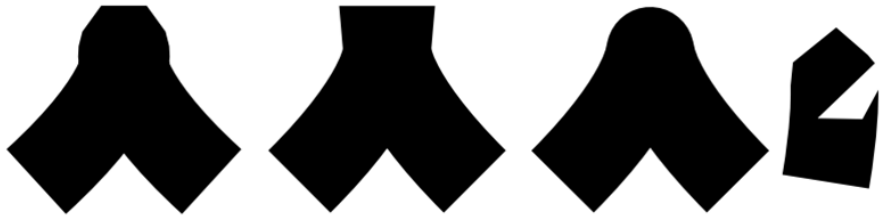

Grade: **C**, two leftmost paths are clipped at top, while rightmost path is noticeably faceted.

## 4.7 GSview 5.0, GhostScript 9.16

Details: GSview 5.0, GhostScript 9.16, circa 2015.

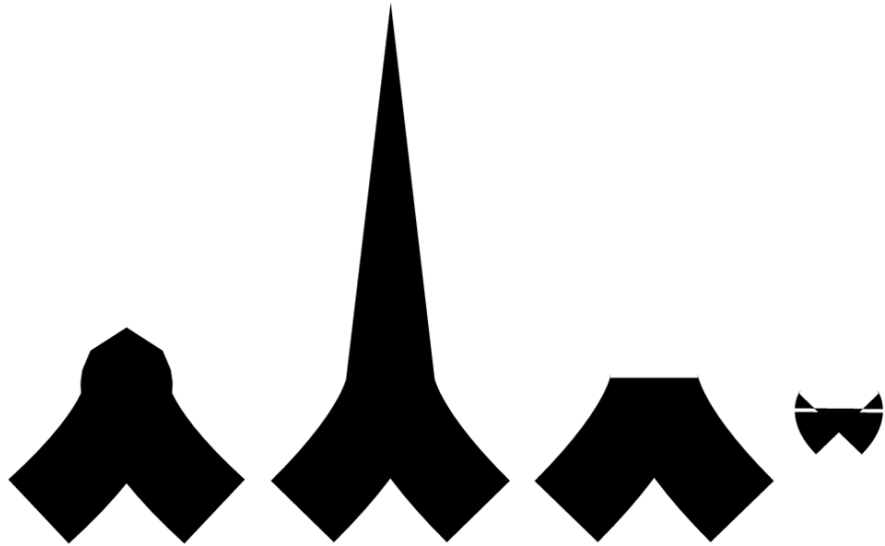

Grade: **D−**, all wrong with the near exact cusp obviously over extended.

## 4.8 Illustrator CC 2019

Details: Adobe Illustrator CC 22.1, 2019.

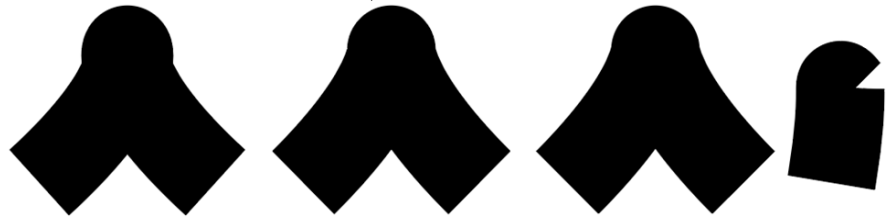

Grade: **A+**.

## 4.9 Inkscape 0.91

Details: Inkscape 0.91, circa 2015.

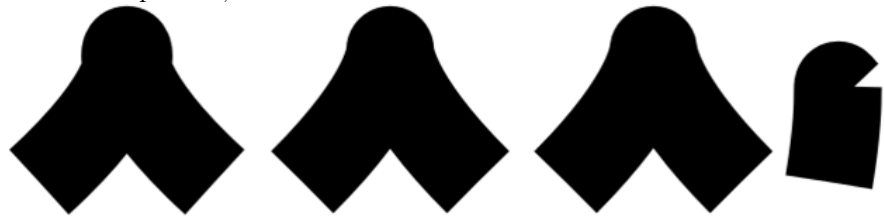

Grade: **A+**.

## 4.10 Internet Explorer 11

Details: Internet Explorer 11.1106, released 2019.

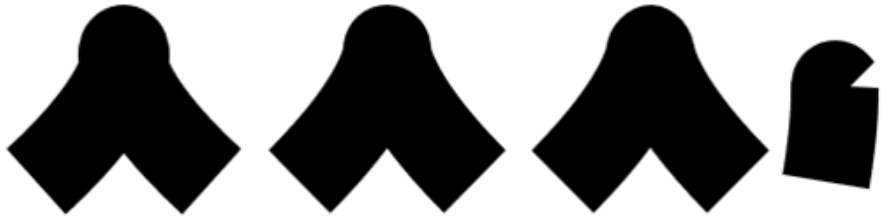

Grade: **A+**, likely due to Internet Explorer 11 relying on Direct2D.

### 4.11 Microsoft Expression Design 4

Details: Microsoft Expression Design 4, version 8.0.31217.1, circa 2012. Loading the Windows Enhanced Metafile version.

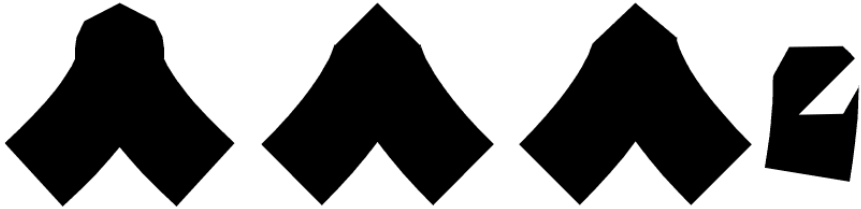

Grade: **D**, all faceted.

### 4.12 Microsoft Office Picture Manager

Details: Microsoft Office Picture Manager (11.8161.1.8405) SP3. Component of Office 2003. Loading the Windows Enhanced Metafile version.

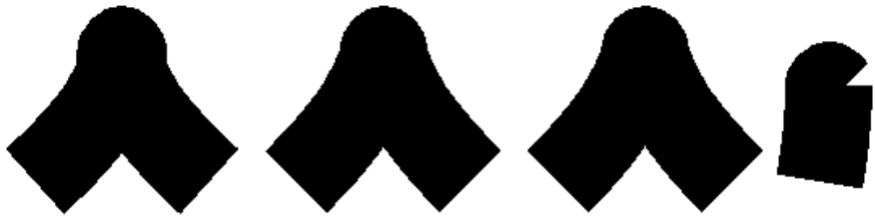

Grade: **A−**, nice except for the ragged antialiasing.

### 4.13 NV_path_rendering

Details: NV_path_rendering, NVIDIA-specific OpenGL extension for path rendering, circa 2015.

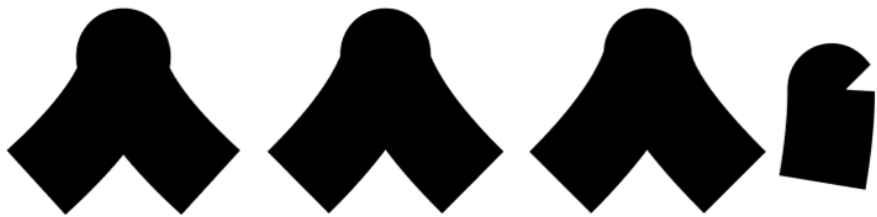

Grade: **A+**.

## 4.14 OpenVG 1.1 Reference Implementation

Details: OpenVG 1.1 Reference Implementation, circa 2008.

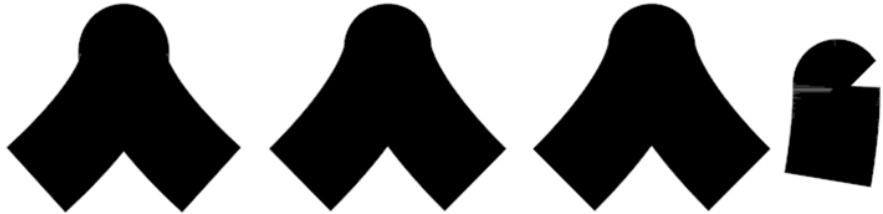

Grade: **A−**, fine stroke shape but there is horizontal and (more subtle) vertical grayness on the fourth case.

## 4.15 Paint Shop Pro 7

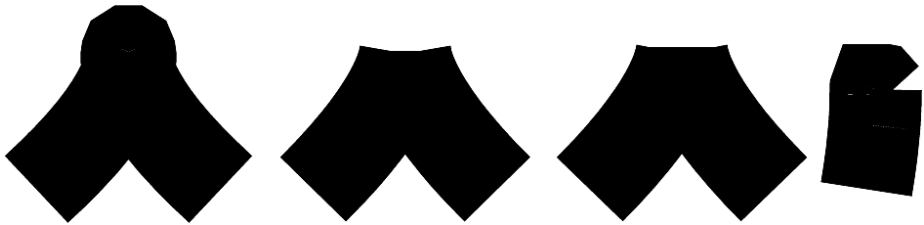

Grade: **D**, all cases faceted.

## 4.16 Qt 4.5

Details: Qt 4.5, circa 2009.

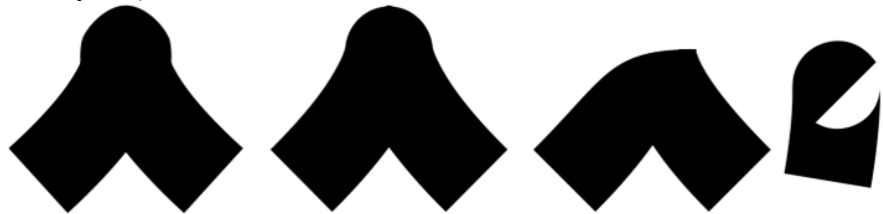

Grade: **C+**, the second case has a small top dimple; two right-side paths have a misshapen cusp and a rounded notch.

## 4.17 Skia

Skia is Google's 2D graphics library used by both Android 2D rendering and Chrome browser.

### 4.17.1 Skia CPU

Details: Skia 2D graphics library, circa 2015.

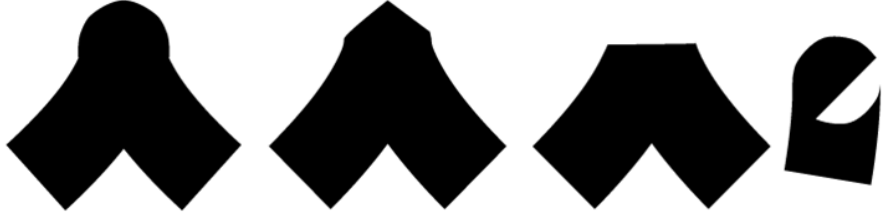

Grade: **C**, middle two paths have faceted cusps while rightmost path is a circular notch.

### 4.17.2 Skia without NV_path_rendering

Details: Skia 2D graphics library, using NVIDIA GPU but without using NV_path_rendering for rendering, circa 2015.

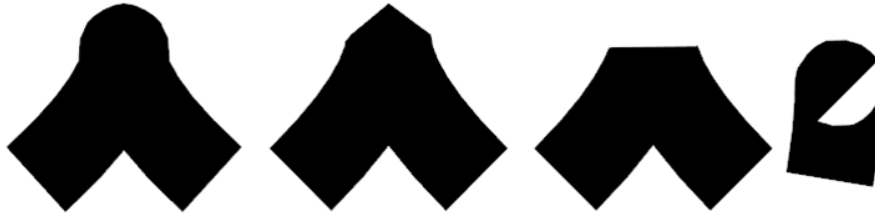

Grade: **C**, same as Skia CPU.

When Skia supports NV_path_rendering, it gets an **A+**, looking essentially the same as Section 4.13.

## 4.18 SumatraPDF 3.1

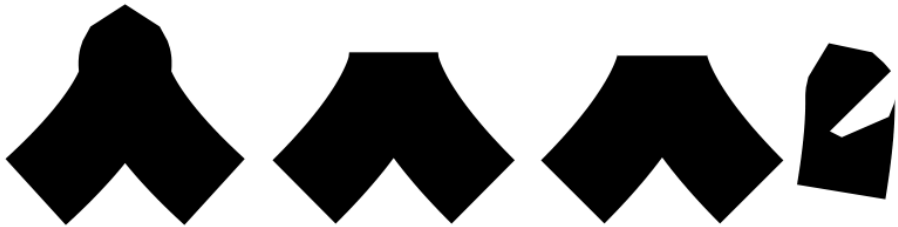

Grade: **D**, leftmost path is has faceted near cusp; middle two paths are chopped off; and rightmost path is faceted including the notch.

### 4.19 Your Current PDF Viewer or Printer

Lastly in the spirit of *you* personally joining our survey, we used the XeTeX PSTricks macro package to embed the four test cases into the PDF you are reading (or you printed).

You can rate for yourself the four test cases:

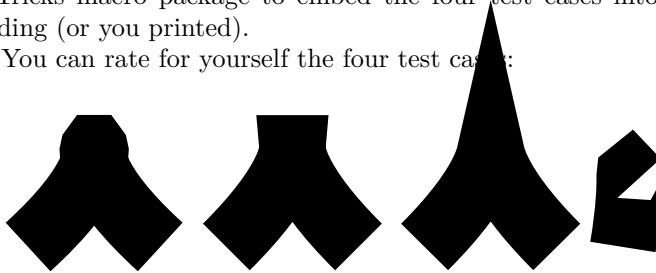

How did your PDF reader or printer do? We observe:

- Adobe Acrobat Reader DC, version 2019.021.20058, received an **A+**.

- Chrome 79 received a **C** (only the third exact cusp case was correct).

- Firefox 72.0.1 received an **A+**.

- Foxit Reader, version 9.7.1.29511, received a **D−** (overflowing the text case into the paper's text above).

- Internet Explorer 72.0.1 received an **A+**.

- Preview on a Mac running OS X 10.4.5 or 10.9.4 received a **C** by having two holes in the left-middle path along with a malformed cap; the right-middle path is horizontally clipped, and the rightmost path has a circular notch.

- SumatraPDF 3.1 received a **D−** (overflowing the text case into the paper's text above).

- TeXworks 0.6.3 (the software used to write this paper), relying on the Poppler PDF rendering library, received a **D**.

## 5 Grading Discussion

We admit these grades are harsh in that they focus entirely on what are arguably obscure corner cases in how paths are stroked; still our intent was to shine a bright light on long-known corner cases.

Eight of our software implementations received **A+** grades by meeting all four criteria *and* being essentially mutually indistinguishable. We contend this is reasonable evidence for the existence of a tacit consensus as to what a stroked path should look like despite vector graphics standards providing no mathematically grounded description of stroking.

For all the software implementations graded a **B** or worse, there was little consistency across different implementations and often at least one (often several) of the test cases was noticeably unappealing when compared to the **A+** graded results.

We made a cursory check of path filling for the same software implementations based on the same four test case paths, just filled instead of stroked. We identified no rendering variances. This makes sense given that path filling has a rigorous, mathematically grounded specification that all implementations appear to implement consistently.

# 6 Conclusion

It has been almost 30 years since Corthout and Pol observed PostScript's stroke operator (as implemented at the time) "has not been implemented robustly" based on a set of four test cases they provided. Our admittedly anecdotal survey demonstrates much variability still exists in how real-world software implements path stroking.

Our survey provides evidence for a consensus based on recent software implementations from Adobe, the developers of Inkscape, Microsoft, and NVIDIA about how a stroked path should be rasterized. This consensus matches our survey's grading criteria.

A more rigorous, mathematically grounded specification of what pixels are rasterized when a path is stroked [6] should reduce the variability among software implementations and give creators of vector graphics content more confidence in the interoperability of the stroke-based content they generate.

# Epilogue

Dutch typographer and writing theorist Gerrit Noordzij lays out a more philosophical argument, reading like gnostic poetry, for knowing the shape of a stroke:

> *The theory furthers good practice too.*
> *The stroke is the fundamental artefact.*
> *Nothing goes further back than the shape of a single stroke.*
> *We cannot postpone a shape by drawing outlines first, because any*
>    *drawing (outlines included) begins with a shape.*
> *Outlines are the bounds of shapes.*
> *If there is not yet a shape, there is no outline either.* [7]